\documentclass[twocolumn,showpacs,aps,prb,superscriptaddress,amsmath,amssymb]{revtex4}
\usepackage{graphicx}
\usepackage{color}
\usepackage[squaren]{SIunits}
\usepackage{dsfont}

\begin{document}
\title{Monitoring Entanglement Evolution and Collective Quantum
  Dynamics}

\author{Georg M. Reuther} \email{georg.reuther@physik.uni-augsburg.de}
\affiliation{Institut f\"ur Physik, Universit\"at Augsburg,
  Universit\"atsstra{\ss}e~1, D-86135 Augsburg, Germany}

\author{David Zueco} \affiliation{Instituto de Ciencia de Materiales
  de Arag{\'o}n y Departamento de F\'{\i}sica de la Materia
  Condensada, CSIC-Universidad de Zaragoza, 50009 Zaragoza, E-Spain}

\author{Peter H\"anggi} \affiliation{Institut f\"ur Physik,
  Universit\"at Augsburg, Universit\"atsstra{\ss}e~1, D-86135
  Augsburg, Germany}

\author{Sigmund Kohler} \affiliation{Instituto de Ciencia de
  Materiales de Madrid, CSIC, Cantoblanco, E-28049 Madrid, Spain}
\date{\today}

\begin{abstract}
  We generalize a recently developed scheme for monitoring coherent
  quantum dynamics with good time-resolution and low backaction
  [Reuther \textit{et al.}, Phys.\ Rev.\ Lett.\ \textbf{102}, 033602
  (2009)] to the case of more complex quantum dynamics of one or
  several qubits.  The underlying idea is to measure with lock-in
  techniques the response of the quantum system to a high-frequency ac
  field. We demonstrate that this scheme also allows one to observe
  quantum dynamics with many frequency scales, such as that of a qubit
  undergoing Landau-Zener transitions. Moreover, we propose how to
  measure the entanglement between two qubits as well as the
  collective dynamics of qubit arrays.
\end{abstract}

\pacs{ 42.50.Dv, 
  03.65.Yz, 
  03.67.Lx, 
  85.25.Cp 
}

\keywords{perturbation theory, quantum computing, quantum measurement,
  quantum electrodynamics, superconducting cavity resonators,
  superconducting integrated circuits}

\maketitle

\section{Introduction\label{sec:intro}}
A most fascinating question in quantum mechanics is how much a
measurement acts back on a quantum system and thus influences the
state of the latter.~\cite{WallsMilburn1995a,Nielsen2000a,Breuer2001a}
According to the classic measurement postulate, the projective
measurement of a quantity causes a collapse of the wave function into
an eigenstate of the corresponding operator. However, quantum
mechanics only allows probabilistic statements about the state into
which the wave function collapses. This evolution into a mixture can
be modeled by coupling the quantum system via the measurement operator
to a measurement apparatus, i.e., to a macroscopic environment which
provokes dissipation and decoherence. A natural interpretation of this
process is that information about the quantum system leaks into the
environment, while the system acquires entropy.
Since the environment is perceived as classical object, one may assume
that the pointer of a measurement apparatus corresponds to a
collective environment coordinate.~\cite{Zurek2003a} Then the question
arises how the quantum system and this collective coordinate influence
each other. In other words, which information about the quantum system
is contained in the collective coordinate and how strong is the
backaction of the environment on the system?

A specific example for such a measurement device is a low-frequency
``tank circuit'' coupled to a superconducting
qubit.~\cite{Grajcar2004a,Sillanpaa2005a} Here, one makes use of the
fact that the resonance frequency of the oscillator depends on the
qubit state which, in turn, influences the phase of the oscillator
response. This allows one to measure both the charge and the flux
degree of freedom of superconducting qubits. The drawback of this
scheme, however, is that the coherent qubit dynamics is considerably
faster than the driving. Thus, one can only observe the time-average
of the qubit state, but not time-resolve its dynamics. Measuring the
qubit directly by driving it at resonance is possible as
well.~\cite{Greenberg2005a} This however induces Rabi oscillations
and, thus, alters the qubit dynamics
significantly.~\cite{Liu2006a,Wilson2007a}.

In Ref.~\onlinecite{Reuther2009a}, we proposed to probe a quantum
system with a weak high-frequency drive acting directly upon the
qubit. We found that the reflected signal possesses a time-dependent
phase shift which is related to a qubit observable. This relation was
validated for a Cooper-pair box that is in a superposition of the two
lowest energy eigenstates which form a qubit. In particular, it was
shown that coherent oscillations of this qubit are visible in the
reflected signal, while the driving-induced backaction stays at a
tolerable level. This enables time-resolved monitoring of the coherent
qubit dynamics.

In this article we demonstrate that this measurement scheme is
applicable also to more complex quantum dynamics, and even the
evolution of entanglement between two qubits can be observed in this
way. In Sec.~\ref{sec:monitoring}, we introduce the underlying
system-bath model and describe how we quantify measurement fidelity
and backaction. Moreover, we derive a relation between the reflected
ac signal and the monitored time-dependent expectation value of the
quantum system. This relation, which constitutes our measurement
scheme, is numerically tested in Sec.~\ref{sec:LZ} for the case of a
qubit undergoing Landau-Zener sweeps. In
Sec.~\ref{sec:entanglement}, we present a protocol for monitoring
entanglement between two qubits. Finally, we derive in
Sec.~\ref{sec:collective} how the collective dynamics of a qubit array
enhances the signal.


\section{High-frequency response of a quantum system}
\label{sec:monitoring}

\subsection{System-bath model\label{sec:SB}}

\begin{figure}[t]
  \includegraphics[width=.8\linewidth]{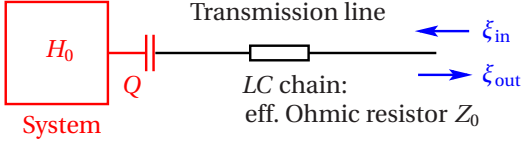}
  \caption{ (Color online) Sketch of the input-output formalism: A
    quantum system described by a Hamiltonian $H_0$ is coupled via its
    charge degree of freedom $Q$ to an environment in form of a
    transmission line with ohmic effective impedance $Z_0$. When the
    system is probed by an input signal $\xi_\mathrm{in}$, the output
    signal $\xi_\mathrm{out}$ contains information about the bare
    system dynamics.  }\label{fig:input-output}
\end{figure}

We consider a quantum system interacting with a dissipative
environment, as depicted in Fig.~\ref{fig:input-output}. The
system-bath Hamiltonian is given
by~\cite{Leggett1987a,Hanggi1990a,Makhlin2001a}
\begin{equation}\label{eq:sb}
  H = H_0 
  +\sum_k\Big(\frac{\varphi_k^2}{2L_k}+\frac{(q_k
    -\lambda_kQ/\omega_k)^2}{2C_k}\Big)\; .
\end{equation}
Here, $H_0$ denotes the system Hamiltonian and $Q$ is the system
operator that couples to the environment. To be specific, we assume
that $Q$ is the system excess charge. In the realm of circuit QED, the
bath is typically modeled by a transmission line in terms of $LC$
circuits with charges $q_k$ and conjugate fluxes $\varphi_k$, where
$C_k$ and $L_k$ are effective capacitances and inductances,
respectively. Furthermore, $\omega_k=(L_kC_k)^{-1/2}$ denotes 
the angular frequency of mode $q_k$, and $\lambda_k$ are the
corresponding coupling constants in units of frequencies.  The
system-bath interaction is fully characterized by the spectral density
\begin{equation}
  \label{eq:spectraldens}
  I(\omega) = \frac{\pi}{2} \sum_k \lambda_k^2
  \sqrt{\frac{L_k}{C_k}}\,
  \delta(\omega-\omega_k) \; ,
\end{equation}
which is assumed to be ohmic with an effective impedance $Z_0$,
i.e. $I(\omega) = \omega
Z_0$.~\cite{Yurke1984a,Devoret1995a,Makhlin2001a}
Assuming a weak system-bath interaction, we obtain the Bloch-Redfield
master equation for the reduced system density operator $\rho$ by
standard techniques,~\cite{FonsecaRomero2005a,Grifoni1998a}
\begin{equation}
  \label{eq:master}
  \dot\rho = \mathcal L_0 \rho = 
  -\frac{i}{\hbar}[H_0,\rho] 
  - \frac{1}{\hbar}
  [ Q, [\hat Q, \rho] ]
  - i\frac{Z_0}{\hbar} [ Q, [ \dot Q, \rho ]_+ ] ,
\end{equation}
where $[A,B]_+ = AB+BA$ denotes the anticommutator, $\dot Q =
i[H_0,Q]/\hbar$ and
\begin{equation}
  \hat Q = \frac{1}{\pi} \int_0^\infty d\tau \int_0^\infty d\omega\,
  S(\omega) \cos(\omega\tau) \tilde Q(-\tau) .
\end{equation}
Here, $S(\omega) = I(\omega)\coth(\hbar\omega/2k_{\rm B} T)$ denotes
the Fourier transformed of the symmetrically ordered equilibrium
bath correlation function at temperature $T$. The shorthand notation
$\tilde X(t)$ represents the interaction-picture operator
$U_0^\dagger(t) X U_0(t)$, where $U_0$ denotes the system propagator.

\subsection{Input-output formalism\label{sec:inout}}

%
%
The central concept of our measurement scheme is to relate the quantum
dynamics of the externally probed central circuit to a response via
the transmission line. In order to quantify the response, we employ
the input-output formalism.~\cite{Gardiner1985b,Johansson2006a} We
start from the Heisenberg equation of motion (operator label $H$) for
the environmental mode $k$,
\begin{equation}
  \label{eq:bathmodes-eq}
  \ddot q_k^H  +\omega_k^2 q_k ^H = \lambda_k \omega_k Q ^H \; ,
\end{equation}
which is linear in the bath operators. Its 
formal solution for initial time $t_0=0$ reads
\begin{multline}
  \label{eq:bathmodes-sol}
  q_k ^H (t) = q_k  (0) \cos (\omega_k t) 
  +  \sqrt{\frac{C_k}{L_k}} \varphi_k (0) \sin (\omega_k t) \\
  + \lambda_k \int_0^t d\tau \sin [\omega_k (t-\tau)] Q^H (\tau) \; .
\end{multline}
It depends on the history of the system, which gets expressed through
the convolution integral in the second line of
Eq.~\eqref{eq:bathmodes-sol}. Inserting the obtained solution into the
Heisenberg equation of motion for any system observable $X$
yields
\begin{equation}\label{eq:sysop-eq}
\begin{split}
  \dot X ^H = & \frac{i}{\hbar} \left[ H_0^H, X ^H \right]
  +\frac{i}{\hbar} \left[ \left (Q^H \right)^2, X ^H \right] 
  \sum_k  \sqrt{\frac{L_k}{C_k}} \frac{\lambda_k ^2}{2\omega_k}\\
  &- \frac{i}{\hbar}
  \left[ Q^H, X^H\right] \sum_k \lambda_k ^2
  \sqrt{\frac{L_k}{C_k}}
  \\ & \times
  \int_0^t d \tau \sin[\omega_k(t-\tau)] Q ^H(\tau) 
  - \frac{i}{\hbar} [Q^H,X^H]
  \\ & \times
  \sum_k \lambda_k \left(
  \sqrt{\frac{L_k}{C_k}}
  q_k (0) \cos (\omega_k t) +  \varphi_k (0)
    \sin  (\omega_k t) \right) .
\end{split}
\end{equation}
At this point we define the operator
\begin{equation}
  \label{eq:xi-in}
  \xi_\mathrm{in}^\mathrm{qm} (t) = \sum_k \lambda_k
  \left( \sqrt{\frac{L_k}{C_k}}
    q_k (0) \cos (\omega_k t) + \varphi_k (0)
    \sin (\omega_k t) \right) \; ,
\end{equation}
which allows a convenient notation. It is fully determined by the
bath correlation function and, following the above equation of
motion, Eq.~\eqref{eq:sysop-eq}, only depends on the environment
operators at initial time. Therefore, it can be interpreted as a
signal entering via the transmission line, i.e.\ describes the input
noise acting upon the system.

After a partial integration of the third term of
Eq.~\eqref{eq:sysop-eq}, the counter-term proportional to
$[(Q^H)^2,X^H]$ is canceled. Furthermore, for the ohmic spectral
density $I(\omega) = \omega Z_0$, Eq.~\eqref{eq:sysop-eq} becomes
local in time, and one arrives at the quantum Langevin
equation~\cite{Ford1988a,Hanggi2005a,Hanggi2009a}
\begin{multline}\label{eq:langevin-in}
  \dot X^H = \frac{i}{\hbar} \left[ H_0^H, X^H\right]
  - \frac{iZ_0}{\hbar}  \left[ Q^H, X^H \right] \dot Q^H  \\
  - \frac{i}{\hbar} [Q^H,X^H ] \xi_\mathrm{in}^\mathrm{qm}(t) \; .
\end{multline}
It contains the input field $\xi_\text{in}^\mathrm{qm}(t)$ and an
ohmic dissipative term $\propto Z_0\dot Q^H$.

It is also possible to express the quantum Langevin equation in terms
of the outgoing fluctuations. This can be
achieved by solving Eq.~\eqref{eq:bathmodes-eq} for the bath modes
starting at a later time $t_1>t$. The solution is the time-reversed
counterpart of Eq.~\eqref{eq:bathmodes-sol} and reads
\begin{multline}
  \label{eq:bathmodes-sol-II}
  q_k^H (t) = q_k (t_1) \cos [\omega_k (t-t_1)]
  +  \sqrt{\frac{C_k}{L_k}} \varphi_k (t_1) \sin[\omega_k (t-t_1)] \\
  - \lambda_k \int_t^{t_1} d\tau \sin [\omega_k (t-\tau)] Q^H(\tau) \; .
\end{multline}
According to Eq.~\eqref{eq:xi-in}, we define the sum over all modes
$k$ with respect to the first two terms of
Eq.~\eqref{eq:bathmodes-sol-II} as the outgoing noise
$\xi_\mathrm{out}^\mathrm{qm}(t)$. In contrast to  
$\xi_\text{in}^\mathrm{qm}(t)$, it is influenced by the system
at earlier times $t<t_1$ and, thus, contains information about the
system dynamics.~\cite{Gardiner1985b} Proceeding as above, we obtain
the time-reversed Langevin equation
\begin{multline}\label{eq:langevin-out}
  \dot X ^H = \frac{i}{\hbar} \left[ H_0^H, X^H \right] \\
  + \frac{iZ_0}{\hbar} 
  \left[Q^H , X^H \right] \dot Q^H - \frac{i}{\hbar} [Q^H,X^H]
  \xi_\mathrm{out}^\mathrm{qm}(t)  \; .
\end{multline}
Its characteristic is a negative damping term, i.e., it is formally
obtained from the quantum Langevin equation~\eqref{eq:langevin-in} via
the substitution $Z_0 \rightarrow - Z_0$. This corresponds to backward
propagation in time, as Eq.~\eqref{eq:langevin-out} only accounts for
the environment at later times $t_1>t$.

The difference between the two Langevin equations relates the input to
the output fluctuations as~\cite{Gardiner1985b}
\begin{equation}\label{eq:in-out}
  \xi_\text{out}^\mathrm{qm}(t)  - \xi_\text{in}^\mathrm{qm}(t)
  = -2 Z_0 \dot Q^H 
  = -\frac{2i Z_0}{\hbar}\big[H_0^H,Q^H\big] .
\end{equation}
The last equality has been obtained from the Heisenberg equation of
the charge operator, $\dot Q ^H = (i/\hbar) [H_0^H,Q^H]$,
which is independent of the environment since $Q$ commutes with the
system-bath coupling Hamiltonian. Equation~\eqref{eq:in-out}
determines the influence of the system on its environment and, thus,
represents the central relation of the input-output formalism.
At this stage, it is worth emphasizing the environment
possesses many degrees of freedom which all acquire information
about the quantum system.  Therefore, the collective bath coordinates
$\xi_\text{in/out}^\mathrm{qm}(t)$ are classical in the sense that
their expectation values can be interpreted as the outcome of a single
measurement.~\cite{Zurek2003a}

\subsection{Response to high-frequency driving \label{sec:highrf}}
The central idea of our scheme is to excite the system by a classical
ac drive and to measure the resulting system response.  Physically,
the driving enters like the quantum noise via the transmission line.
Therefore, it can be modelled as a coherently and highly excited
bath mode in the classical limit weakly coupled to the
system.~\cite{Grifoni1998a}
This means that the input fluctuations are augmented by a
deterministic ac contribution. Thus, they become 
$\xi_\mathrm{in} (t) = \xi_\mathrm{in}^\mathrm{qm}(t) + A\cos(\Omega
t)$ and, accordingly, $\xi_\mathrm{out} (t) =
\xi_\mathrm{out}^\mathrm{qm}(t) + A \cos (\Omega t)$. The
deterministic terms also affects the input-output relation 
\eqref{eq:in-out} which now reads
\begin{equation}
\label{eq:in-out-total}
  \xi_\text{out} (t)  = \xi_\mathrm{in}^\mathrm{qm} (t) 
  + A\cos(\Omega t)  -\frac{2i Z_0}{\hbar} [H_0^H,Q^H] \, .
\end{equation}
In the corresponding expectation value, where we return
to the Schr\"odinger picture for convenience,
the fluctuation  $\xi_\mathrm{in}^\mathrm{qm}$ vanishes, such that,
\begin{equation}\label{eq:in-out-exp}
  \langle\xi_\text{out} (t)\rangle = A\cos(\Omega t)
   -\frac{2i Z_0}{\hbar}\langle[H_0,Q]\rangle_t \,.
\end{equation}
This expresses the expectation value of the output field in terms of
the input field and a system expectation value,
$\langle\ldots\rangle_{t} = \mathrm{tr}[\rho (t)\ldots]$. The latter
contains information about the system dynamics and is the central
quantity of interest. 

The quantum Langevin equations~\eqref{eq:langevin-in} and
\eqref{eq:langevin-out} have been convenient for
deriving the relations between the input and the output fields,
Eqs.~\eqref{eq:in-out} and \eqref{eq:in-out-total}.  For
the actual computation of the system expectation value on the r.h.s.\
of Eq.~\eqref{eq:in-out-exp}, by contrast, such stochastic,
operator-valued equations are less practical.  Therefore, we derive
in addition an equivalent quantum master equation which not only
serves for numerical computations, but will also provide an analytical
high-frequency approximation.

We start by noticing that a classical ac drive $A\cos(\Omega t)$
coupled to the system charge $Q$ corresponds to a Hamiltonian $QA\cos(\Omega
t)$.  Thus, the system Hamiltonian $H_0$ has to be replaced by
 \begin{equation}\label{eq:H0-drive}
   H(t) \equiv H_0 + QA\cos(\Omega t) \; .
 \end{equation}
A key requirement of our measurement scheme is that the driving must
not significantly alter the system dynamics. Thus, we assume that the
amplitude $A$ is sufficiently small, which allows us to treat the
driving perturbatively.~\cite{Kohler1997a} In doing so, we use the
ansatz $\rho(t) = \rho_0(t) + \rho_1 (t)$, where $\rho_{0}(t)$
describes the dynamics of the system without ac driving,
and $\rho_{1}(t)$ denotes the correction to the system state.  With
this ansatz, the master equation reads
\begin{equation}\label{eq:perturb-rho}
  \dot \rho(t) = \big[\mathcal L_0 + \mathcal L_1 (t) \big]
\big[\rho_{0}(t) + \rho_{1}(t)\big] \; ,
\end{equation}
where the perturbative driving manifests in the Liouvillian
\begin{equation}\label{eq:qme-drive}
  \mathcal L_1 (t) \rho (t) = - \frac{i}{\hbar} A 
  [Q,\rho (t)] \cos \Omega t \, ,
\end{equation}
To lowest order in $A$, the correction $\rho_1$ obeys
\begin{equation}
  \label{eq:perturb-rho1}
  \dot\rho_1 (t) = \mathcal{L}_0\rho_1 (t) -\frac{i}{\hbar}
  A[Q,\rho_0 (t)]\cos(\Omega t) \; .
\end{equation}
This linear inhomogeneous equation of motion can be solved formally in
terms of a convolution between the propagator of the undriven system
and the inhomogeneity,
\begin{equation}
  \label{eq:rho1}
  \rho_1 (t) =  e^{\mathcal{L}_0 t}\rho_1 (0) -\frac{i}{\hbar}
  \int\limits_0^{\;t} d \tau \, e^{\mathcal{L}_0(t-\tau)} A
  [Q,\rho_0(\tau)] \cos(\Omega \tau) \; ,
\end{equation}
with $\rho_1 (0) =0$.
If the driving frequency $\Omega$ is much larger than all relevant
system frequencies, the integral can be simplified by time-scale
separation: First, we split the above integral into a sum of integrals
over complete periods and a final one over the remaining
time. Assuming that the slow $\rho_0(t)$ is practically constant
during one oscillation period of $2\pi/\Omega$, the integrals over
complete driving periods vanish, while the last contribution can be
evaluated using $\rho_0 (\tau) \approx \rho_0 (t)$. In this way, we
obtain the solution for the time evolution of the reduced density
matrix
\begin{equation}
  \label{eq:rho}
  \rho(t) = \rho_0(t) - \frac{iA}{\hbar\Omega}
  [Q,\rho_0 (t)] \sin(\Omega t) \; , 
\end{equation}
where $eA/\hbar \Omega$ is identified as the necessarily small
perturbation parameter. Within this approximation for the
density operator, the expectation value of the output,
Eq.~\eqref{eq:in-out-exp}, becomes
\begin{multline}\label{eq:xiout1}
  \langle\xi_\text{out}(t)\rangle = A \cos(\Omega t) -\frac{2i
    Z_0}{\hbar}\langle[H_0,Q]\rangle_{0,t} \\
  + \frac{2 A Z_0}{\hbar^2\Omega} \left\langle [[H_0,Q],Q ]
  \right\rangle_{0,t} \sin(\Omega t) \; ,
\end{multline}
where the subscript in $\langle\ldots\rangle_{0,t} =
\mathrm{tr}[\rho_0(t)\ldots]$ refers to the undriven dynamics.

At this point we note that $\langle\xi_\text{out}(t)\rangle$ contains
both low-frequency components stemming from the pure system dynamics
as well as high-frequency components induced by the external driving.
The latter corresponds to the second term of Eq.~\eqref{eq:rho}. Now, in an
experiment, it is feasible to single out the high-frequency components
of the outgoing signal using a lock-in technique, where the incoming
signal represents the reference oscillator. This removes the second
term on the r.h.s.\ of Eq.~\eqref{eq:xiout1}, such that the outgoing
signal becomes
\begin{equation}
  \label{eq:xiout}
  \langle\xi_\text{out}^\mathrm{hf}(t)\rangle
  = A \cos(\Omega t) + \frac{2 A Z_0}{\hbar^2\Omega} 
  \left\langle [[H_0,Q],Q ] \right\rangle_{0,t} \sin(\Omega t) .
\end{equation}
Since the second term is a perturbative correction, it has to be small
so that Eq.~\eqref{eq:xiout} can be written as
\begin{equation}
  \label{eq:xiout-phase}
  \langle\xi_\text{out}^\mathrm{hf}(t)\rangle = A
  \cos[\Omega t - \phi_\mathrm{hf}^0(t)] 
\end{equation}
with the time-dependent phase shift
\begin{equation}\label{eq:phase}
  \phi_\mathrm{hf}^0(t)
  = \frac{2 Z_0}{\hbar^2\Omega}
  \left\langle [[H_0,Q],Q ] \right\rangle_{0,t} \; .
\end{equation}
This relation constitutes the basis of our measurement scheme. It
connects a phase shift $\phi ^0_\mathrm{hf}(t)$ between the
high-frequency input and the output signals to the low-frequency
dynamics of a system observable. In other words, Eq.~\eqref{eq:phase}
enables \textit{time-resolved} monitoring of an open quantum system by
measuring the phase shift $\phi^0_\mathrm{hf}(t)$ with lock-in
techniques. We emphasize that our measurement scheme is rather generic
and can in principle be applied to any open quantum system.

It is important to note that there are situations in which the phase
shift $\phi_\mathrm{hf}^0(t)$ is independent of the system dynamics or
even vanishes. The latter is obviously the case when $Q$ commutes with
the system Hamiltonian. A constant phase shift is obtained for an
harmonic oscillator $H_0 = \hbar\omega_0 (a^\dagger a + 1/2)$ that
couples via its position, i.e., $Q\propto a + a^\dagger$, as one
easily obtains by evaluating the double commutator in
Eq.~\eqref{eq:phase}. This is evident from the superposition principle
which holds for a linearly driven harmonic oscillator.  However,
already a coupling $Q\propto (a+a^\dagger)^2$ is sufficient for
obtaining a non-trivial response.

\subsection{Measurement fidelity and backaction\label{sec:fid}}

As broached above, the relation between the phase of the output signal
and a system observable relies on a high-frequency approximation. In
an experiment, the driving frequency is finite, however. Thus, the
experimentally obtained phase $\phi_\text{out}^\text{exp}(t)$, which
is actually recorded by the lock-in amplifier, may differ from the
theoretically predicted phase $\phi ^0_\text{hf}(t) \propto \langle
[[H_0,Q],Q] \rangle_{0,t}$. During lock-in amplifying, the
high-frequency components of $\langle \xi_\text{out} ^\text{hf}(t)
\rangle$, and with it $\phi_\text{out}^\text{exp}(t)$, are extracted
from the output signal $\langle \xi_\text{out}(t) \rangle $, where the
classical incoming signal $\langle \xi_\text{in}(t) \rangle =
A\cos(\Omega t)$ serves as the reference oscillator.

Thus, it is crucial to test with a numerical
simulation how well both phases agree in a realistic case. As a
criterion, we employ the measurement fidelity which we define as the
normalized overlap 
\begin{align}
  F =& \Big(\phi_\text{out}^\text{exp} ,\phi_\text{hf}^0 \Big)
  \nonumber\\
  \equiv& \left(\int dt\, [\phi_\text{out} ^\text{exp}(t)]^2 \int dt\,
    [\phi^0_\text{hf}(t)]^2\right)^{-1/2}
  \nonumber \\ & \times
  \Big|\int dt\,
  \phi^\text{exp}_\text{out}(t) \, \phi ^0_\text{hf}(t) \Big| \; ,
  \label{eq:fidelity}
\end{align}
with time integration over the duration of the measurement. The ideal
value $F=1$ corresponds to the case of perfect proportionality between
the measured phase $\phi_\mathrm{out}^\text{exp}(t)$ and
$\phi_\text{hf}^0 (t) \propto \langle[[H_0,Q],Q] \rangle_{0,t}$. Note
that $\phi_\text{hf}^0 (t)$ is computed in the absence of the ac
driving, while $\phi_\text{out}^\text{exp}(t)$ is obtained in
its  presence. Moreover, we have to mimic the action of the lock-in
amplifier numerically, as described in Sec.~\ref{sec:LZ}
in the second paragraph after Eq.~\eqref{eq:Qdot}.

Furthermore, for any quantum measurement, one has to worry about
backaction on the system in terms of decoherence. In our measurement
scheme, decoherence plays a particular role, because both the driving
and the ohmic environment couple to the central system via the same
mechanism. This is reflected by the fact that the predicted phase of
Eq.~\eqref{eq:phase} is proportional to the dissipation strength $Z_0
$. In terms of the generalized, dimensionless dissipation strength
$\alpha = e^2 Z_0 /\hbar$, it is required that $\alpha \lesssim 0.1$
in order to preserve a predominantly coherent time evolution. In this
context, our measurement is weak, but yet destructive in the sense
that it relies on the naturally provided interaction with a
dissipative environment causing decoherence. This marks a significant
difference to conventional quantum non-demolition (QND) measurement
schemes.~\cite{WallsMilburn1995a,Breuer2001a} Here by contrast, the
additional decoherence due to the driving is minor. This is obvious
from the fact that within first-order approximation the system purity
is not affected,
\begin{equation}
  \mathrm{Tr}(\rho^2) = \mathrm{Tr} (\rho_0^2) \, ,
  \label{eq:trsq}
\end{equation}
which follows readily from Eq.~\eqref{eq:rho}. Only in this sense, our
measurement scheme can be considered as being of non-demolition
character.

As a concrete measure for how much the driving perturbs the system, we
use in our numerical investigations the time average $\bar D$ of the
trace distance~\cite{Nielsen2000a}
\begin{equation}
  D(t) = \frac{1}{2} \mathrm{Tr}|\rho(t)-\rho_0(t)| \; .
  \label{eq:trdist}
\end{equation}
In the ideal case, $\bar D$ vanishes, while $\bar D=1$ if both density
operators are completely unrelated.

\section{Monitoring Landau-Zener sweeps\label{sec:LZ}}
In Ref.~\onlinecite{Reuther2009a}, we investigated the quality of our
measurement scheme using the dynamics of a decaying qubit state as an
elementary example. Here, we go a step further and demonstrate that it
is applicable to more complex coherent dynamics.

As an example, we consider a Landau-Zener tunneling process between
the states of a charge qubit prepared in a Cooper-pair box
(CPB),~\cite{Wendin2006a, Shevchenko2010a} a paradigmatic example of
complex quantum dynamics in a seemingly simple system. In particular,
it can be used for state preparation~\cite{Saito2006a,Zueco2010a} and
entanglement generation~\cite{Wubs2007a,Zueco2008a} in a
qubit-resonator system. In recent experiments~\cite{Oliver2005a,
  Sillanpaa2006a} multiple Landau-Zener sweeps were performed with a
charge qubit coupled to a low-frequency tank oscillator whose
resonance frequency depends on the time-averaged qubit state. This
enabled the measurement of the accumulated phase which determines the
qubit state. Here by contrast, we demonstrate that a time-resolved
observation of the complex dynamics during a single Landau-Zener
transition is possible as well.

We employ the CPB Hamiltonian
\begin{equation}
  \label{eq:CPB}
  H_0^\mathrm{CPB}
  = 4 E_\mathrm{C}(\hat N-N_g)^2 -\frac{\delta}{2}
  \sum_{N=-\infty}^\infty (|N{+}1\rangle\langle N|+\text{h.c.}) ,
\end{equation}
where $N$ is the number of excess Cooper pairs in the box. The charge
operator reads $Q=2e\hat N = 2e\sum_N N|N\rangle\langle N|$ with $e$
being the elementary charge. The charging energy $E_\mathrm{C}$ is
determined by the various capacitances of the CPB, while the scaled
gate voltage $N_g$ and the effective Josephson energy $\delta$ are
controllable.
If the charging energy is sufficiently large, and $N_g$
approaches a charge degeneracy point determined by the half-integer
value $N_g^{\mathrm{deg}} \approx N_0-1/2$, for some integer number
$N_0$, only the two lowest charge states $|N_0-1\rangle \equiv
|{\downarrow}\rangle$ and $|N_0\rangle\equiv |{\uparrow}\rangle$
matter and form a qubit, while the energy gap to the higher charge
states is much larger than the qubit splitting.~\cite{Makhlin2001a,
Vion2002a, Grajcar2004a} In the two-level approximation, the qubit is
described by the Hamiltonian
\begin{equation}\label{eq:qb}
  H_0^\mathrm{qb} =
  -\frac{1}{2}\epsilon\sigma_z -\frac{1}{2}\delta\sigma_x \, .
\end{equation}
The Pauli matrices $\sigma_i$ are defined in the qubit subspace and
$\epsilon = 8E_\mathrm{C}(N_0-N_g)$, where $N_g$ ranges in the
interval $[N_0-1,N_0]$, is the
effective qubit bias. Moreover, $Q_\mathrm{qb} = e\sigma_z$ while by
virtue of relation~\eqref{eq:phase} the phase of the output signal is
linked to the qubit observable $\sigma_x$ according to
\begin{equation}\label{eq:phaseangle}
  \phi_\mathrm{hf}^0(t)
  = - \frac{ 4e^2 Z_0 \delta}{\hbar^2 \Omega}
  \left\langle \sigma_x  \right\rangle_{0,t}\, .
\end{equation}
Thus, the high-frequency component of $\langle\dot Q\rangle$, which is
manifest in the phase of the outgoing signal \eqref{eq:in-out},
contains information about the low-frequency qubit dynamics in terms
of the unperturbed $\langle\sigma_x\rangle_0$. It is worth mentioning
that the preceding discussion equally applies to the case of a flux or
phase qubit with an inductive coupling to the transmission line.

The qubit undergoes a Landau-Zener transition when the gate voltage in
terms of the background charge $N_g$ in the Hamiltonian~\eqref{eq:CPB}
is swept from a large positive to a negative finite value through the
charge degeneracy point at $N_g^{\mathrm{deg}}$ with constant
velocity, $N_g(t) = N_g^0 + v_g t$. Accordingly, the qubit
Hamiltonian~\eqref{eq:qb} depends on time and can be written as
\begin{equation}
  H_0(t) = -\frac{v(t-t_0)}{2} \sigma_z -\frac{\delta}{2} \sigma_x \, ,
  \label{eq:qblz}
\end{equation}
where we have defined $v = 8E_C v_g$ and
$t_0=-N_g^0/v_g$. Hamiltonian~\eqref{eq:qblz} is a valid approximation
on condition that the sweep ends timely prior to a subsequent
anti-crossing in the CPB spectrum.
%
\begin{figure}[tb]
  \includegraphics[scale=.95]{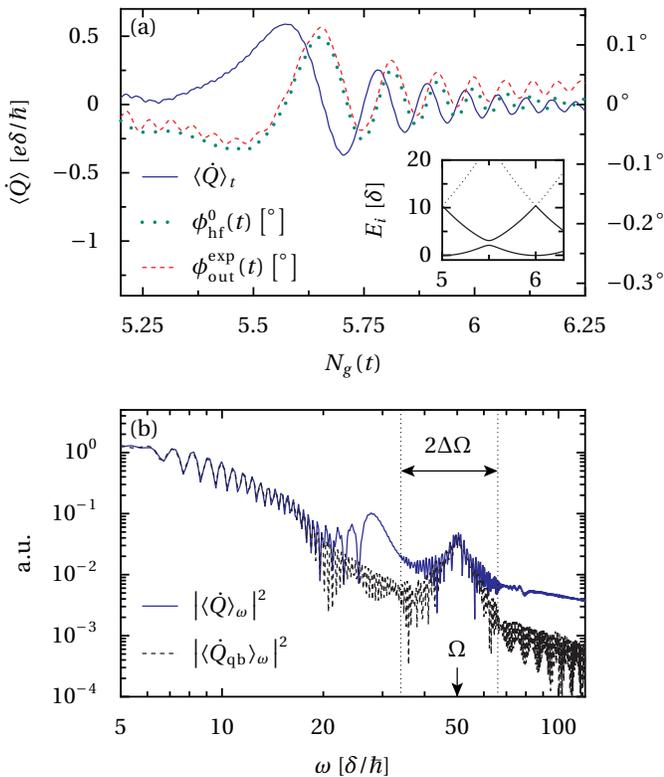}
\caption{(Color online)
Monitoring the dynamics of a dissipative Landau-Zener sweep of a
qubit, which is initiated in a weakly probed CPB with $N_\mathrm{s}
=12$ states operated at $N_0=6$. The selected anticrossing lies
at $N_g = N_0-1/2= 5.5$, and the initial qubit state is
$|{N=5}\rangle \equiv |{\uparrow}\rangle$, while $|{\downarrow}\rangle
\equiv |{N=6}\rangle$. The other parameters are $E_c = 2.6\,\delta$,
$v_g = 0.125\,\delta/\hbar$, $A=\delta/e$, 
$\alpha=0.05$, and $\Omega = 50\,\delta/\hbar$.
(a) Time evolution of the output signal $\langle \dot Q \rangle_t
\propto \langle\xi_\mathrm{out} (t)\rangle  -
\langle\xi_\mathrm{in}(t)\rangle$ 
of the full CPB, and its
lock-in amplified phase $\phi_\text{out}^\text{exp} (t)$ (frequency
window $\Delta\Omega=16 \delta$), compared to the estimated phase
$\phi_\mathrm{hf}^0 (t) \propto \langle\sigma_x \rangle_{0,t}$ in the
qubit approximation. 
Fast oscillations with frequency $\Omega=50\delta/\hbar$ stemming
from the input signal are barely resolved. The inset shows the three
lowest CPB eigenenergies as a function of $N_g(t)$.
(b) Power spectrum of $ \langle\dot Q  \rangle$ for the driven
systems, comparing the full CPB Hamiltonian (solid) and in two-level
approximation (dashed).
}\label{fig:LZ}
\end{figure}

Even if the qubit approximation describes the CPB faithfully for large
charging energies, $E_c \gg \delta$, one must take into account
excitations of higher charge states by the ingoing signal; see also
the discussions in Ref.~\onlinecite{Reuther2009a}. In our
simulations, we operate at $N_0=6$ and truncate the CPB Hilbert space
at $N_\mathrm{s}=12$ states, which turned out to be sufficient to
reach numerical convergence.

Figure~\ref{fig:LZ}(a) illustrates the time evolution of the CPB
observable
\begin{equation}
  \label{eq:Qdot}
  \dot Q  = \frac{i e \delta}{\hbar} 
  \sum_{N=-\infty}^\infty (|N\rangle\langle N+1|-|N+1\rangle\langle N|)
  \; ,
\end{equation}
during a Landau-Zener transition between the two lowest states; see
inset of Fig.~\ref{fig:LZ}(a).  Within two-level approximation, this
operator reads $\dot Q = (-e \delta /\hbar) \sigma_y$.
The parameters are chosen such that the final populations of the two
states are roughly equal, because in this crossover regime between the
adiabatic and the non-adiabatic dynamics, coherent oscillations are
most pronounced.  Note that since we plot the time derivative of the
charge, $\langle \dot Q \rangle \propto \langle \sigma_y \rangle$, the
final population of the upper state, $ [1 + \langle \sigma_z \rangle
(t\rightarrow \infty )]/2$, given by the Landau-Zener probability
$P_\mathrm{LZ} =  \exp(-\pi\delta^2/2\hbar v) \approx 0.546$, is not 
part of the information depicted in Fig.~\ref{fig:LZ}(a).

The backaction of the weak driving signal on the system
dynamics in terms of a weak modulation of $\langle\dot Q \rangle_t$ is barely 
visible. The associated spectrum in terms of $\langle\dot Q
\rangle_\omega$, depicted in panel (b), nevertheless reflects the
driving in terms of a small peak at the driving frequency
$\Omega=50 \delta/\hbar$. Due to the time-dependent qubit splitting,
the spectrum exhibits a broad range of frequencies centered at zero,
reflected in the sideband structure around $\Omega$. In the
time domain these sidebands correspond to a signal 
$\langle\xi_\mathrm{out}^\mathrm{hf} (t)\rangle = A \cos[\Omega t
-\phi_\mathrm{out}^\text{exp} (t)]$.

As is indicated above in Secs.~\ref{sec:highrf} and~\ref{sec:fid}, the
phase $\phi_\mathrm{out}^\text{exp}(t)$ can be retrieved by lock-in
amplification of the output signal in an experiment. We mimic this
procedure numerically in the following way:~\cite{Scofield1994a} We
only consider the spectrum of $\langle \xi_\text{out} (t)\rangle$ in a 
frequency window $\Omega\pm\Delta\Omega$ around the driving frequency
and shift it by $-\Omega$ in order to center it at zero. The inverse
Fourier transformation of the spectrum cutout to the time domain
yields the time-dependent phase $\phi_\mathrm{out}^\text{exp}(t)$ which is
expected to agree with $\phi_\text{hf}^0(t)$; see the discussion in
Sec.~\ref{sec:fid}. According to Eq.~\eqref{eq:phase}, this phase
reflects the unperturbed time evolution of $\langle\sigma_x\rangle_0$
with respect to the qubit. 

  Figure~\ref{fig:LZ}(a) shows that the numerically extracted phase
  $\phi_\mathrm{out}^\text{exp} (t)$ is indeed in a very good agreement 
  with $\phi^0_\text{hf} (t) \propto \langle\sigma_x\rangle_{0,t}$ for
  an appropriate choice of parameters. This holds even if the
  condition of high-frequency probing, i.e. $\Omega$ exceeding all
  relevant system frequencies by far, is not strictly
  fulfilled. Appropriate values for $\Omega$ need to be determined from
  the width of the sideband distributions. These can be estimated from
  the duration $\tau_\mathrm{LZ}$ of a Landau-Zener transition which
  is $\tau_\mathrm{LZ} \approx 20 \hbar / \delta$ here. Consequently,
  the spectral peak has the width $v\tau_\text{LZ}/\hbar \approx 20
  \delta / \hbar$. Thus, in order to avoid any overlap between both
  distributions, the driving frequency must fulfill $\Omega\gg 20
  \delta / \hbar $. In the present case, Eqs.~\eqref{eq:xiout}
  and~\eqref{eq:phase} are corroborated already for $\Omega \approx 50
  \delta /\hbar$.

Furthermore, we require small dissipation, $\alpha \lesssim 0.1$, to
keep the time evolution predominantly coherent. In this context, one
needs to take care of the interplay of decoherence and the external
sweep velocity and its impact on the LZ probability and
dynamics.~\cite{Wubs2006a, Zueco2008a, Nalbach2009a}

Towards the end of the sweep, one notices a deviation between the
predicted phase $\phi ^0_\text{hf} (t)$ and
$\phi_\text{out}^\text{exp}(t)$ which becomes curved upward. This
stems from the presence of higher charge states, which get slightly
populated as soon as $N_g$ reaches the next half-integer value
$N_g^{\mathrm{deg}}+1$, that is, the subsequent avoided level
anticrossing in the CPB spectrum. Accordingly, as
visible in Fig.~\ref{fig:LZ}(b), the spectrum $\langle\dot
Q\rangle_\omega$ contains more pronounced frequency components than
the one of the two-level approximation $\langle\dot
Q_\mathrm{qb}\rangle_\omega$.  However, the deviations between both
spectra are minor, which corresponds to the good agreement between
$\phi_\mathrm{out}^\text{exp}(t)$ and $\phi_\mathrm{hf}^0(t)$.
\begin{figure}[tb]
  \includegraphics{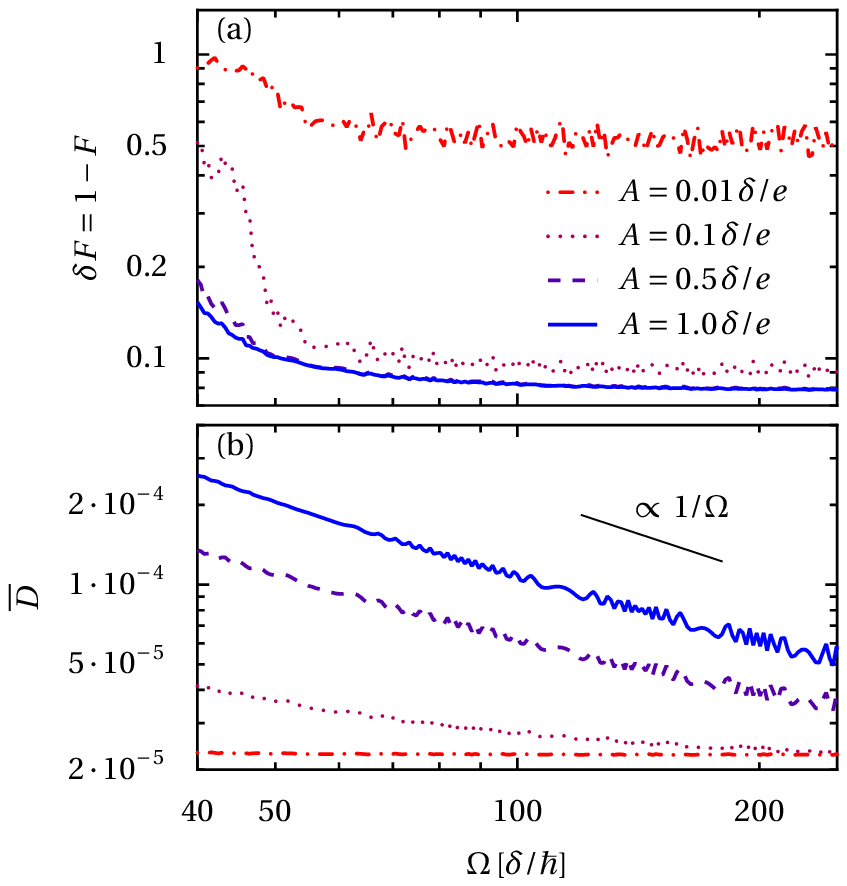}
  \caption{(Color online) (a) Fidelity defect $\delta F=1-F$ and (b)
    time-averaged trace distance $\overline D$ between the density
    operators of a driven and an undriven CPB during a LZ
    sweep. Both quantities are depicted for various driving amplitudes
    $A$ as a function of the driving frequency $\Omega$. All other
    parameters are the same as in Fig.~\ref{fig:LZ}.  }
  \label{fig:fidelity}
\end{figure}

The measurement fidelity defect $\delta F = 1-F$ between
$\phi_\mathrm{out}^\text{exp}(t)$ and $\phi^0_\text{hf}(t)$, computed with
Eq.~\eqref{eq:fidelity}, is depicted in Fig.~\ref{fig:fidelity}(a) as
a function of the driving frequency. Apparently, resonant excitations
to non-qubit CPB states play a minor role here since, in contrast to
the case of a time-independent qubit energy,~\cite{Reuther2009a} we
do not observe characteristic resonance peaks in $\delta F$. Since all
CPB frequencies are varied in time during the LZ sweep, the population of
non-qubit states by the driving is mainly suppressed. Furthermore, the
fidelity only slightly increases after the driving frequency $\Omega$
has exceeded a minimal value of $v \tau_\mathrm{LZ}/\hbar$. This
corroborates that $\Omega \approx 50 \delta/\hbar$ is a good choice.

With regard to the driving amplitude, one has to find a compromise.
The phase contrast of the outgoing signal~\eqref{eq:xiout1} obviously
increases with $A$, whereas the driving perturbs more and more the
low-frequency dynamics. Here, the choice of $A = \delta/e$ yields a
maximum fidelity of $F\simeq 0.9$ for the above driving frequency.
This apparently low value essentially stems from the discussed
deviation between $\phi_\text{out}^\text{exp} (t)$ and $\phi
^0_\text{hf}(t)$. Far from a subsequent LZ sweep, the system is
faithfully described with only the qubit levels. All in all, for
these values, $e A/\hbar\Omega \approx 0.01$--$0.1$, which
justifies our perturbative treatment.

The small-dissipation condition $\alpha\lesssim 0.1$ together with the
above conditions on the driving amplitude and frequency provides phase
shifts of roughly $\phi_\mathrm{out}^\text{exp} \approx 0.1^{\circ}$,
which is small but still measurable with present technologies. In
particular, the constraint of large driving frequencies $\Omega$ for
monitoring a LZ sweep reduces the maximum visibility of the
oscillations encoded in the phase shift. In contrast, a smaller value
of $\Omega$, if applicable, may result in a visibility of the order of
$1^{\circ}$. We turn back to these quantitative issues in
Sec.~\ref{sec:collective} where we discuss the collective dynamics of
equal quantum systems.
 
Generally, decoherence may be influenced by ac
driving.~\cite{Kohler1997a} Here, however, we do not observe a
significant change of decoherence. We conclude this from
Fig.~\ref{fig:fidelity}(b) which depicts  the time-averaged trace
distance, computed with Eq.~\eqref{eq:trdist} as a function of the
driving frequency for various driving amplitudes. For intermediate
amplitudes, we find roughly $\overline D\propto A/\Omega$, unless $A$
is very small, which agrees with relation~\eqref{eq:trsq}. This states
that at high frequencies, the driving does not add decoherence, which
confirms the picture drawn by investigating the measurement fidelity
$F$. 

Our model Hamiltonian does not consider the excitation of
quasi-particles in the superconductor. Therefore it is valid only as
long as the driving frequency stays below the gap energy. Thus, a
setup made of aluminum the driving is limited to $\Omega \lesssim
100\,\mathrm{GHz}$. With a typical Josephson energy of the order of
some GHz, the driving frequency required to monitor the dynamics of a
LZ sweep already comes close to this limit. An implementation with
niobium, whose energy gap is considerably larger~\cite{Novotny1975a}
and which has been used in circuit QED experiments,~\cite{Berns2008a}
should be less critical.

\section{Two-qubit entanglement\label{sec:entanglement}}

When considering a second qubit, it may be entangled with the first
one. In the present context, this raises the question whether our
measurement scheme is sensitive to this property. In order to dwell
into this question, we focus on a system of two charge qubits
described by the Hamiltonian
\begin{equation}
  \label{eq:twoqubit}
  H_\mathrm{2q,0} = \frac{\omega}{2} 
  (\sigma_z^{(1)} + \sigma_z^{(2)}) + g \sigma_x^{(1)} \sigma_x^{(2)} ,
\end{equation}
where the upper indices label the qubits. Here, we assume that both
qubits possess the same energy splitting $\omega$ and are mutually
coupled with strength $g$. A qubit coupling scheme as described here
is effectively realized in the case of, e.g., two CPBs  within
two-level approximation which are capacitively coupled to a
transmission line resonator. In analogy to the above
  discussions, our measurement scheme can be applied on condition that
  the driving field is not in resonance with transitions to higher
  energy levels of the underlying physical
  system.~\cite{Reuther2009a} If the 
qubit-resonator interaction is dispersive, i.e., if the resonator is
far detuned with respect to the qubits, it mediates an effective
qubit-qubit $XX$-coupling,~\cite{Zueco2009b} such that the two-qubit
system can be described by the effective
Hamiltonian~\eqref{eq:twoqubit}. Alternatively, an
  effective $XX$-coupling between two qubits can be mediated by a
  third, dispersively detuned qubit.~\cite{Niskanen2007a}

Both charge qubits couple capacitively to a common
environment via their charge operator
\begin{equation}
  \label{eq:twoqubit-sb}
  Q_\mathrm{2q} = e (\sigma_z^{(1)} + \sigma_z^{(2)}) \, .
\end{equation}
An adequate measurement of the qubit-qubit entanglement is the
concurrence
\begin{equation}
  C=\max\{\chi_1-\chi_2-\chi_3-\chi_4,0\} \; ,
  \label{eq:concurrence-gen}
\end{equation}
where the $\chi_j$ denote the ordered square roots of the eigenvalues
of the matrix $\rho (\sigma_{y}^{(1)} \sigma_{y}^{(2)}) \rho^*
(\sigma_{y}^{(1)} \sigma_{y}^{(2)})$ with $\rho$ denoting the
two-qubit density matrix.~\cite{Wootters1998a}

For later convenience we express the system operators in terms of the
four, maximally entangled Bell states,
\begin{align}
  \label{eq:bellstates}
  |\Phi_\pm\rangle &= \frac{1}{\sqrt{2}} \left( |00\rangle \pm |11
    \rangle
  \right) \; , \\
  |\Psi_\pm\rangle &= \frac{1}{\sqrt{2}} \left( |10\rangle \pm |01
    \rangle \right) \, .
\end{align}
In the basis
$\{|\Phi_+\rangle,|\Phi_-\rangle,|\Psi_+\rangle,|\Psi_-\rangle\}$ the
two-qubit Hamiltonian~\eqref{eq:twoqubit} reads
\begin{equation}
  \label{eq:twoqubit-sb1}
  H_\mathrm{2q,0} = \frac{1}{2} 
  \begin{pmatrix}
    \hphantom{-} g & - \omega & \,0 & 0 \\
    -\omega & -g & \,0 & 0 \\[.5ex]
    0 & 0 & \,g & 0 \\
    0 & 0 & \,0 & -g
  \end{pmatrix} ,
\end{equation}
while the charge operator becomes
\begin{equation}
  \label{eq:twoqubit-sb-2}
  Q_\mathrm{2q} = 2 e
  \begin{pmatrix}
    0 & 1 & 0 & 0 \\
    1 & 0 & 0 & 0 \\
    0 & 0 & 0 & 0 \\
    0 & 0 & 0 & 0
  \end{pmatrix} .
\end{equation}
This representation of the coupling to the environment evidences that
neither the 
system-bath coupling nor the ac driving generate transitions between
the subspaces spanned by the states $\{|\Phi_\pm\rangle\}$ and
$\{|\Psi_\pm\rangle\}$, respectively. Concerning the latter states, it
is even such that they are not coupled to the environment at all,
i.e., they form a decoherence-free subspace.~\cite{Lidar1998a,Yu2002a}
For this reason, we henceforth focus on initial preparations in the
subspace spanned by $\{|\Phi_\pm\rangle\}$. In this particular case,
the concurrence is given by
\begin{equation}
  \label{eq:concurrence}
  C_{\{|\Phi_\pm\rangle\}} (t)
  = 2 |\rho_{0011} (t)|
  \equiv 2 |\mathrm{Tr} \{\rho(t) |00\rangle \langle 11| \} | ,
\end{equation}
i.e., by an off-diagonal density matrix element in the product basis.
In terms of the pseudo-spin operators $\sigma_z '$ and $\sigma_x'$
defined in the basis $\{|\Phi_+\rangle,|\Phi_-\rangle\}$, 
the Hamiltonian and the charge operator become
\begin{align}
  H_\mathrm{2q,0} ' & = g \sigma_z' - \omega \sigma_x ' \;
  ,  \label{eq:twoqubit-subspace-sys} \\ 
  Q_\mathrm{2q} ' & = 2 e \sigma_x' \; ,
  \label{eq:twoqubit-subspace-int}
\end{align}
respectively.

\begin{figure}[t]
  \includegraphics[width=.9\linewidth]{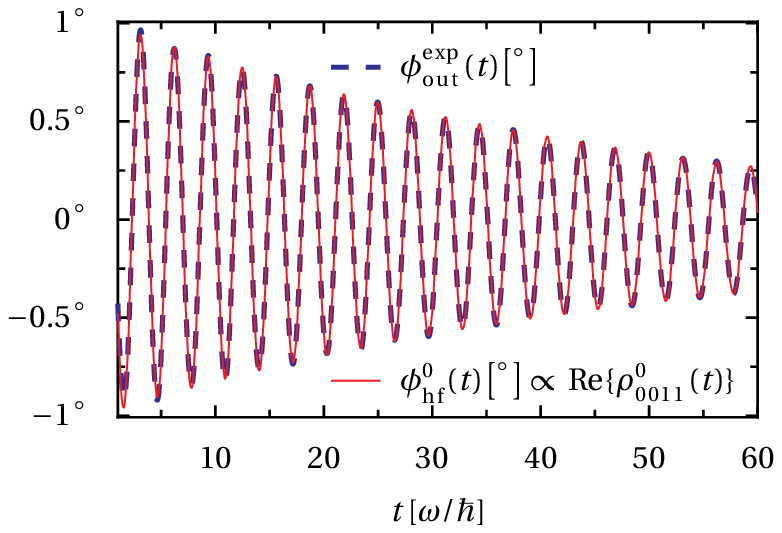}
  \caption{ (Color online) Monitoring the entanglement evolution of
    two coupled qubits with initial state ${|}\Phi_+\rangle$. The
    lock-in amplified phase $\phi_\text{out}^\text{exp}$ (blue dashed line) is
    compared to the estimated phase $\phi _\mathrm{hf}^0(t)$ (red
    solid line) which is proportional to the real part of the density
    matrix element $\rho^0_{0011}(t)$. The latter is needed for
    reconstructing the concurrence $C(t)=2|\rho^0_{0011}(t)|$ of the
    unperturbed system. Parameters are $\Omega = 15\,\omega$,
    $g=0.1\,\omega$, $A = 0.1\,\omega$ and $\alpha =
    0.08$.} \label{fig:ent-phase}
\end{figure}

At this point, we recall Eq.~\eqref{eq:phase} which relates the phase
of the outgoing signal to a particular observable of the undriven
system. Inserting the subspace Hamiltonian and charge operator,
Eqs.~\eqref{eq:twoqubit-subspace-sys} and
\eqref{eq:twoqubit-subspace-int}, we obtain
\begin{equation}\label{eq:phase-ent}
  \phi_\mathrm{hf}^0(t)
  = \frac{32 g \alpha }{\hbar\Omega}
  \left\langle \sigma_z' \right\rangle_{0,t}\; ,
\end{equation}
with the dimensionless damping strength $\alpha = e^2 Z_0 /\hbar$, and
\begin{equation}\label{eq:phase-C}
  \langle \sigma_z' \rangle_{0,t}
  \equiv \mathrm{Tr} \{\rho_0 (t) \sigma_z '\} 
  = 2 \mathrm{Re} \{ \rho_{0011}^0 (t) \} \, .
\end{equation}
In other words, the phase of the outgoing signal is proportional to
the real part of the density matrix element $\rho_{0011}^0 (t)$ of the
unperturbed system. Figure~\ref{fig:ent-phase} demonstrates that this
relation holds sufficiently well: It compares the numerically
calculated phase of the lock-in measurement, $\phi_\mathrm{out}^\text{exp}(t)$,
to the phase $\phi_\mathrm{hf}^0(t)\propto \langle \sigma_z '
\rangle_{0,t}$ predicted by the measurement relation~\eqref{eq:phase}. We 
find an excellent agreement between both quantities, while the maximum
angular visibility reaches values of around $1^\circ$. Thus, we find
that it is possible to access directly the key density matrix element
needed to reconstruct the concurrence $C_{\{|\Phi_\pm\rangle\}} (t)$
for the subspace $\{|\Phi_\pm\rangle\}$.
%
%

Monitoring $\mathrm{Re} \{\rho_{0011}^0 (t) \}$ already provides a
lower bound for the entanglement and may even provide a good estimate
for it. Nevertheless, it is desirable to measure also its imaginary
part. For this purpose one has to rotate the basis of the second qubit
by applying the phase gate
\begin{equation}
  \label{eq:phasegate}
  U_\mathrm{phase}^{(2)} =
  \begin{pmatrix}
    1 & 0\\
    0 & i
  \end{pmatrix} .
\end{equation}
This local unitary transform does not alter the two-qubit entanglement
at the time it is applied. For small values of $g$, it approximately
commutes with the total time evolution operator of the system, $\exp(i
H t/\hbar)$. i.e., $[\exp(i H t/\hbar), U_\mathrm{phase}^{(2)}] =0$
for $g=0$. In analogy to Eq.~\eqref{eq:phase-C}, the relation for the
phase shift then becomes
\begin{equation}\label{eq:phase-C1}
  \phi_\mathrm{hf}^0(t) \propto 
  \mathrm{Im} \{ \rho_{0011}^0 (t) \} \, .
\end{equation}
Thus, we have access to both the real and the imaginary part of the
density matrix element $\rho_{0011}^0 (t)$. One just needs to initiate
two copies of a pure entangled state in the subspace
$\{|\Phi_\pm\rangle\}$, apply the gate of Eq.~\eqref{eq:phasegate} to
one of them, and monitor their dissipative time evolution. The
difference between the two-qubit dynamics in either case due to the
phase gate $U_\mathrm{phase}^{(2)}$ is relevant only at large times
when the qubits approach their stationary state. For the transient
stage depicted in Fig.~\ref{fig:ent-phase}, we have verified that the
phase of the output signal indeed reflects the imaginary part of the
concurrence (not shown). This, in turn, enables one to reconstruct the
characteristics of the concurrence $C_{\{|\Phi_\pm\rangle\}} (t)$ and
thus of the amount of entanglement between the two qubits, according
to Eq.~\eqref{eq:concurrence}.

In a similar way, one finds access to the entanglement when the system
is initially prepared in the subspace $|\Psi_\pm\rangle$. This can be
done by applying the so-called Pauli-X gate to the second qubit,
\begin{equation}
  \label{eq:Xgate}
  U_\mathrm{X}^{(2)} \equiv \sigma_x^{(2)} = 
  \begin{pmatrix}
    0 & 1\\
    1 & 0
  \end{pmatrix} \; ,
\end{equation}
effectively leaving us in the $|\Phi_\pm\rangle$ subspace again, from
where one can pursue as described above. Like this, it is possible to
monitor the time characteristics of entanglement of all kinds of
entangled two-qubit states, as long as the subspaces
$\{|\Phi_\pm\rangle\}$ and $\{|\Psi_\pm\rangle\}$ are not mixed.

Admittedly, the relation between the entanglement and
$\phi_\text{hf}^0$, manifest in Eqs.~\eqref{eq:phase-ent} and
\eqref{eq:phase-C}, holds only if the concurrence can be expressed as
the expectation value of an observable providing the required
off-diagonal matrix element of the density operator. Obviously, this
does not always apply. However, two qubits prepared in a Bell state
represent such a case which is of relevance and interest.

So far, we have addressed initially pure entangled states, which
constitutes an idealization and is not always achievable in an
experiment. A more realistic scenario copes with the drawback that one
starts from a mixed state which naturally possesses less
entanglement. Moreover, the concurrence is then no longer determined
by a single off-diagonal matrix element [cf.\
Eq.~\eqref{eq:concurrence}], but rather has to be computed from the
definition of the two-qubit concurrence,
Eq.~\eqref{eq:concurrence-gen}. In order to test how much information
about the concurrence $C(t)$ is nevertheless contained in the
measurement signal \eqref{eq:phase-ent}, we study the time evolution
for the preparation in an unpolarized mixture or Werner
state,~\cite{Werner1989a}
\begin{multline}
  \label{eq:mixture}
  \rho_\mathrm{W} = (1-p) |\Phi_+\rangle \langle \Phi_+| \\
  + \frac{p}{3} \left( |\Phi_-\rangle\langle\Phi_-| +
    |\Psi_+\rangle\langle\Psi_+| + |\Psi_-\rangle\langle\Psi_-|
  \right) \; ,
\end{multline}
where $p$ denotes the degree of depolarization. A perfect initial
preparation in the fully entangled state $|\Phi_+\rangle$ is then
characterized by $p=0$. For $p>0.5$, the state is separable. In the
realm of circuit QED, a polarization degree of $p=0.1$ has already
been achieved,~\cite{Ansmann2009a} while in a quantum optical context,
even the much lower value $p\lesssim 0.01$ could be
reached.~\cite{Lougovski2005a}

\begin{figure}[t]
  \includegraphics[width=.9\linewidth]{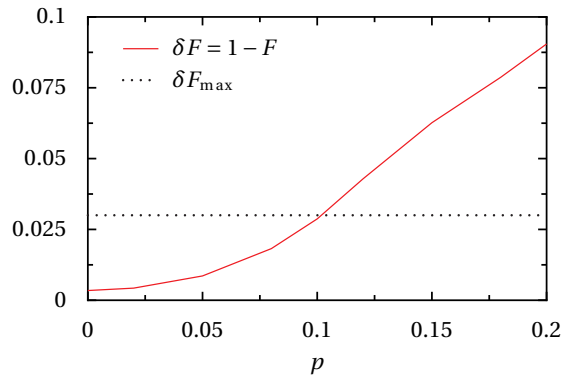}
  \caption{(Color online) Fidelity defect $\delta F = 1 -F$ of
    the concurrence curves $C_{\{|\Phi_\pm\rangle\}} (t)$ and $C(t)$
    in dependence of the degree of depolarization, $p$. The function
    overlap is integrated over the decay time of the two-qubit
    state. As discussed in the text, the upper bound $\delta
    F_\mathrm{max} = 0.03$ marks the largest acceptable value of
    $p$. All other parameters as in Fig.~\ref{fig:ent-phase}.  }
  \label{fig:fidC}
\end{figure}

Our goal is now to compare the concurrence $C(t)$ with
$C_{\{|\Phi_\pm\rangle\}} (t)$ which holds for preparation in a Bell
state. As a measure for the agreement, we use again the fidelity $F$
defined as the scaled overlap between both quantities, i.e., by
Eq.~\eqref{eq:fidelity} but with the phases replaced by the
concurrences. Figure~\ref{fig:fidC} shows the resulting fidelity
defect $\delta F = 1-F$ as a function of the initial degree of
depolarization. From a numerical comparison of the
  concurrence curves  $C(t)$ and $C_{\{|\Phi_\pm\rangle\}} (t)$ for
  different values of $\delta F$ (not shown), we set the threshold for
  the maximum acceptable fidelity defect to $\delta F_\mathrm{max} =
  0.03$. The investigation of several examples showed that, beyond a
  fidelity defect of this order, both curves no longer appear similar.
 Thus, it appears that
  the measured phase shift captures the concurrence dynamics
  satisfactorily only for a $\delta F$ below this
  threshold. Remarkably, the latter is not reached until an initial
  depolarization of $p \lesssim 0.1$. From this, we conclude that our
  measurement protocol allows one to gain information about
  qubit-qubit entanglement in circuit QED even for depolarized initial
  states.

\section{Collective quantum dynamics\label{sec:collective}}
The typical signal that can be reached with our protocol, as regards
measurements of single quantum systems, is a phase shift of the order
$1^{\circ}$, as estimated in Sec.~\ref{sec:LZ}. Of practical interest
is whether this already marks the maximum reachable. In the following,
we suggest how to increase the visibility by monitoring the collective
dynamics of a large number of identical quantum objects at the same
time. For this purpose, we exemplarily investigate an ``atomic cloud''
containing a large number of atoms as described in
Ref.~\onlinecite{Verdu2009a}. There, an ensemble of cold atoms
interacting with a superconducting cavity, described by the Dicke
model, was investigated. Provided that it is possible to single out
one particular atomic transition, the cloud represents an ensemble of
identical two level atoms. This gives rise to the $N$-qubit
Hamiltonian
\begin{equation}
  \label{eq:cloud-H}
  H_{N\mathrm{q},0} = \epsilon \sum_{j=1}^N \sigma_z^{(j)} + \delta
  \sum_{j=1}^N \sigma_x^{(j)} \, . 
\end{equation}
The interaction between the electromagnetic field in the cavity and
the atoms is assumed dipole-like and can be cast into the operator
$Q_{N\mathrm{q}} (a +  a^\dagger)$. Here, the collective coupling
operator $Q_{N\mathrm{q}}$ is chosen dimensionless and given as 
\begin{equation}
  \label{eq:cloud-sb}
  Q_{N\mathrm{q}} = \sum_{j=1}^N \sigma_z^{(j)} = \sqrt{N} S_z \, ,
\end{equation}
and $S_z = \sum_{j=1}^N \sigma_z^{(j)} /\sqrt{N}$ denotes the
collective spin of the cloud. 

In the case where an open transmission line is employed in the setup
rather than the (closed) cavity, a continuum of modes replaces the
single cavity mode in the modeling. In this sense, the transmission
line can be understood as one-dimensional heat bath. The dipole
interaction between the cloud and the environment becomes
$H_{\mathrm{int}} = Q_{N\mathrm{q}} \sum_k c_k (a_k ^ \dagger +
a_k^{\vphantom{\dagger}})$. Thus, one naturally ends up with the
Caldeira-Legget model of Eq.~\eqref{eq:sb} where, formally, $c_k =
\lambda_k \omega_k/C_k$. 

Assuming an ohmic environment, the phase shift~\eqref{eq:phase} then
becomes
\begin{equation}\label{eq:phase-gen}
  \phi_\mathrm{hf}^0(t)
  = \frac{2 \alpha}{\hbar\Omega}
  \left\langle [[H_0,Q],Q ] \right\rangle_{0,t} \; ,
\end{equation}
where $\alpha$ denotes the dimensionless damping strength,
and $\alpha \delta$ is the coupling strength to the
  environment per atom.
Inserting $H_{N\mathrm{q},0}$ and $Q_{N\mathrm{q}}$ into the measurement
relation~\eqref{eq:phase-gen} yields
\begin{equation}\label{eq:phase-cloud}
  \phi_\mathrm{hf}^0(t)
  = \frac{4 \alpha \delta \sqrt{N}}{\hbar\Omega}
  \left\langle S_x \right\rangle_{0,t}
\end{equation}
with $ S_x = \sum_{j=1}^N \sigma_x^{(j)} /\sqrt{N}$. Thus, the phase
signal scales with $\sqrt{N}$ as compared to Eq.~\eqref{eq:qb}. By
virtue of this relation, monitoring the collective dynamics of $N$
two-level systems via the total spin paves the way towards a higher
angular visibility. 

Since the validity of the measurement relation~\eqref{eq:phase-gen} is
restricted to small angles, the visibility may be enhanced up to the
order $10^{\circ}$. In order to be specific, we evaluate
$\phi_\text{hf}^0$ for the parameters~\cite{Verdu2009a}
$\delta/2\pi\hbar = 6.8 \mathrm{GHz}$, $\alpha = 6\cdot 10^{-5}$,
$\Omega/2\pi = 20\mathrm{GHz}$. This yields an effective single-atom
coupling to the ohmic environment of $\alpha\delta/2\pi\hbar \approx
0.4\mathrm{MHz}$. Despite this very small system-bath interaction
strength per atom, the estimated phase resolution for $N=10^{6}$ atoms
in the cloud reaches $\phi_\mathrm{hf}^0 = 4^\circ-5^\circ$. At the same time,
atomic decoherence is drastically reduced. Under these conditions,
which are experimentally achievable,\cite{Verdu2009a} the real-time
monitoring of an atomic ensemble constitutes a major  improvement in
angular visibility in comparison with the single-qubit case.

\section{Conclusions\label{sec:conclusion}}

We have derived a scheme for monitoring the low-frequency dynamics of
a quantum system, where a particular focus has been put on quantum
circuits. The central idea is to couple the quantum system
capacitively to a high-frequency ac field and to measure the reflected
signal. We have shown that the phase of the latter contains
information about a particular expectation value of the central
system. In an experiment, this information can be extracted with
lock-in techniques. This scheme has some similarities to the qubit
readout via a tank circuit~\cite{Grajcar2004a, Sillanpaa2005a} or a
quantum point contact.~\cite{Ashhab2009b} The essential difference is
that the proposed high-frequency driving enables a time-resolved
measurement.

The underlying relation \eqref{eq:phase} between a particular
expectation value and the phase shift of the reflected signal relies
on a time-scale separation. This constitutes an approximation, which
has to be validated. In so doing, we have
demonstrated that the measurement fidelity is rather good. As an
example, we have studied a qubit undergoing Landau-Zener transitions.
Owing to the time dependent level splitting of the qubit, its dynamics
possesses a broad frequency spectrum. Thus, it represents a more
challenging test case than the coherent oscillations considered in
Ref.~\onlinecite{Reuther2009a}. Also for the present case, the
measurement fidelity turned out to be rather good.

A further relevant issue is the backaction of the measurement on the
system. The relevant backaction is decoherence which here is
unavoidable because the driving and the environmental degrees of freedom
couple to the system via the same mechanism. This is reflected by the
fact that the phase shift being the recorded signal is proportional to
the decoherence rate. Thus a stronger coupling to the driving can be
achieved only at the expense of more decoherence.

Concerning possible applications in solid-state quantum information
processing, the generalization towards systems with two or more qubits
is rather important. For the case of two qubits, we have demonstrated
that it is possible to monitor the time evolution of their
entanglement, provided that one can access the relevant density matrix
elements via the phase shift of the reflected signal.  Even though the
entanglement decays due to the unavoidable decoherence, it is still
possible to monitor an appreciable number of cycles between entangled
and separable states of interacting qubits. Furthermore, our protocol
produces reliable results even in the case where the monitored state
is mixed rather than pure.

Finally, we have also addressed the question of how much stronger the
measurement signal can be if the single qubit is replaced by a whole
array of qubits that undergo the same dynamics. The result is that the
signal scales with the square root of the number of qubits. This even
leaves some room for improving the fidelity and reducing the coupling
to each qubit.  As a consequence, decoherence becomes less relevant,
which is encouraging in particular when attempting a
proof-of-principle realization.


\begin{acknowledgments}
  We thank Jens Siewert and Enrique Solano for fruitful discussions.
  This work has been supported by DFG through the Collaborative
  Research Center SFB 631, project A5, and through the German
  Excellence Initiative via ``Nanosystems Initiative Munich
  (NIM)''. G.R. acknowledges financial support from the NIM seed
  funding program.
  D.Z. acknowledges financial support from MICINN under grant numbers
  FIS2008-01240 and FIS2009-13364-C02-0.
\end{acknowledgments}

\end{document}